\documentclass[showpacs,preprintnumbers,amsmath,amssymb,aps,twocolumn]{revtex4}

\usepackage{amsfonts}
\usepackage{bm}
\usepackage[dvips]{graphicx}
\usepackage{color}
\usepackage{multirow}
\usepackage{epsfig}

\newcommand{\eqnref}[1]{Eqn.~\eqref{#1}}
\newcommand{\figref}[1]{Fig.~\ref{#1}}

\newcommand{\tabref}[1]{Table~\ref{#1}}
\newcommand{\diffd}{\text{d}}

\newcommand{\matrixsymb}[1]{\mathbb{#1}}
\newcommand{\neweqnline}{\nonumber\\}
\newcommand{\punc}[1]{\,#1}

\newcommand{\tr}{\operatorname{Tr}}
\renewcommand{\vec}[1]{\mathbf{#1}}

\newcommand{\qD}{q_{\text{D}}}

\newcommand{\SrTiO}{SrTiO$_{3}$ }

\newcommand{\KTaO}{KTaO$_{3}$ }
\newcommand{\EuTiO}{EuTiO$_{3}$ }
\newcommand{\unit}[1]{\text{#1}}

\begin{document}
\title{Theory of quantum paraelectrics and the metaelectric transition}
\author{G.~J.~Conduit}
\email{gjc29@cam.ac.uk}
\affiliation{Theory of Condensed Matter Group, Department of Physics, 
Cavendish Laboratory, 19 J.J.~Thomson Avenue, Cambridge, CB3 0HE, UK}
\author{B.~D.~Simons}
\affiliation{Theory of Condensed Matter Group, Department of Physics, 
Cavendish Laboratory, 19 J.J.~Thomson Avenue, Cambridge, CB3 0HE, UK}
\date{\today}

\begin{abstract}
We present a microscopic model of the quantum paraelectric-ferroelectric
phase transition with a focus on the influence of coupled fluctuating phonon 
modes. These may drive the continuous phase transition first order through a 
metaelectric transition and furthermore stimulate the emergence of a 
textured phase that preempts the transition. We discuss two further 
consequences of fluctuations, firstly for the heat capacity, and secondly 
we show that the inverse paraelectric susceptibility displays $\chi^{-1}
\sim T^{2}$ quantum critical behavior, and can also adopt
a characteristic minimum with temperature. 
Finally, we discuss the observable consequences of our results.
\end{abstract}

\pacs{77.80.Bh, 05.70.Jk, 64.60.-i, 77.84.Dy}

\maketitle

\section{Introduction}

Ferroelectric materials feature in many modern day electronic devices
including computer memory and capacitors, and are a simple setting for studying
quantum criticality~\cite{07tiy07,08khopcsjhhjck10,09rssdlss03}. In this
paper we focus on the family of displacive ferroelectrics where
the optical lattice modes condense, forming
a structural distortion. Near to quantum criticality excitations can become
highly degenerate and new phases can emerge. Motivated by recent experiments
that signal the emergence of novel quantum critical behavior in
ferroelectrics~\cite{07tiy07}, we explore the possibility that transverse
components of polar fluctuating phonons conspire to drive a first order
displacive metaelectric transition and investigate the implications for the
inverse susceptibility.

The soft-mode optical phonons in ferroelectrics can be well-described by a
bosonic field theory. If the dynamics were not damped by free electrons and
the interactions 
remain short-ranged then the general quantum critical behavior
would adhere to the well-established rules reviewed in
Ref.~\cite{00s01}. However, in ferroelectrics the motion of the atoms in
optical modes leads to the emergence of electric dipoles. A good description
of these long-range dipole forces is essential to properly describe the
ferroelectric transition. The effect of long-range dipolar forces was first
studied by~\citet{71r08} and~\citet{71ks09}. \citet{73af10} found that
anisotropies associated with the dipolar interaction led to a universality
class in the classical ferroelectric. The quantum ferroelectric phase
transition in the mean-field approximation, and its universality class, was
studied by~\citet{03rm01}. However, recent experimental evidence points to
new physics that emerges close to quantum criticality; for example the
coexistence of a quantum paraelectric phase with a quantum ferroelectric
phase in $^{18}$O-exchanged \SrTiO provides strong evidence for a first
order phase transition~\cite{07tiy07}. Additional motivation to study
ferroelectrics arises from the inverse dielectric constant behavior of
\SrTiO which falls at low temperature before increasing as
$\epsilon^{-1}\sim T^{2}$ at intermediate temperatures and rises as
$\epsilon^{-1}\sim T$ at high temperature. One suggestion is that new
phenomena are driven by the coupling of acoustic to optical
phonons~\cite{71ks09,09pcc02,09rssdlss03}. However, inspired by the
ramifications of quantum fluctuations in ferromagnets~\cite{08cs12}, we show
that the transverse coupling of fluctuating phonons can drive a first order
metaelectric transition.

Having realized that fluctuations can cause the emergence of a first order
transition it is natural to search for further phase reconstruction.
Motivated by the development of a textured FFLO phase~\cite{64ff08,65lo03},
and evidence for a textured ferromagnetic state near to the ferromagnetic
first order transition~\cite{05bkr06,08ebc04,09cgs06}, here we search for
the emergence of an analogous textured ferroelectric phase. Finally, to
connect to prevailing experimental methods, we derive an appropriate
expression for the inverse susceptibility that is consistent with recent
experimental results~\cite{80rhb07,06c03,09rssdlss03} over a wide range of
the phase diagram, and demonstrate that the transverse coupling of
fluctuating phonons could cause it to have a characteristic minimum at low
temperature.

\section{Action and mean-field theory}\label{sec:FEFormalism}

\begin{figure}
 \centerline{\resizebox{0.85\linewidth}{!}{\includegraphics{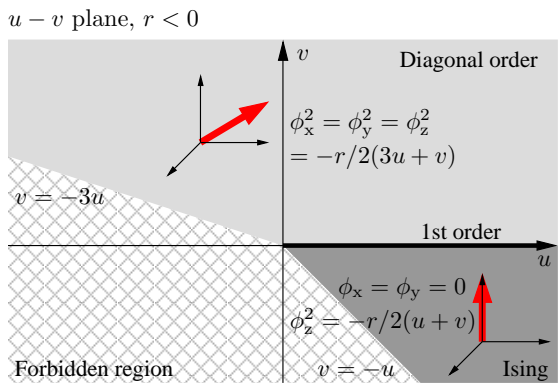}}}
 \caption{(Color online) The phase diagram in the $u-v$ plane
 at zero temperature in the mean-field approximation. The cross-hatched
 forbidden region denotes where the polarizability would diverge without higher
 order corrections. The solid thick line highlights a first order phase boundary
 between the light grey region that denotes diagonal order, and the
 dark grey which labels the Ising phase. In each regime the inset axes
 illustrate the polarization solution highlighted by the red vector.}
 \label{fig:MeanFieldPhaseDiagram}
\end{figure}

We adopt a bosonic field theory to describe the soft optical phonon modes
that should recover the main physical behavior of the system. The order
parameter of the theory is the local polarization $\bm\phi(\vec{x},t)=
\sum_{i=1}^{n}e_{i}\vec{r}_{i}(\vec{x},t)$, which is
formally defined for one unit cell at $\vec{x}$ containing $n$ atoms of
charge $e_{i}$ each individually displaced through $\vec{r}_{i}$ by the
optic mode. As the optical phonon softens, the action develops an
instability and the order parameter must describe both thermal and quantum
fluctuations. Following \citet{03rm01} we describe the action in
three-dimensional space and imaginary time via the Ginzburg-Landau
phenomenology
\begin{eqnarray}
 \!\!\!\!\!\!\!\!&&S=\int_{0}^{\beta}\Biggl\{\sum_{\vec{q},\alpha,\beta}\Biggl[\left(\frac{a^{2}}{c^{2}}\partial_{\tau}^{2}+a^{2}q^{2}+r+fq_{\alpha}^{2}\right)\delta_{\alpha,\beta}\neweqnline
 \!\!\!\!\!\!\!\!&&+\left(g-hq^{2}\right)\frac{q_{\alpha}q_{\beta}}{q^{2}}\Biggr]\phi_{\alpha}(\vec{q})\phi_{\beta}(\vec{-q})\neweqnline
 \!\!\!\!\!\!\!\!&&+\!\!\!\!\!\!\sum_{\alpha,\beta,\{\vec{q}_{i}\}}\!\!\!\!\!\!\left(u+v\delta_{\alpha,\beta}\right)\phi_{\alpha}(\vec{q}_{1})\phi_{\alpha}(\vec{q}_{2})\phi_{\beta}(\vec{q}_{3})\phi_{\beta}(\vec{q}_{4})\Biggr\}\diffd\tau\punc{,}
\end{eqnarray}
where $a$ is the lattice constant, $c$ is the speed of the phonons,
$q^{2}=\sum_{\alpha}q_{\alpha}^{2}$, the dimensionless momenta
$-\pi<q_{\alpha}\le\pi$, and the second summation is carried out under the
conservation of momentum
$(\vec{q}_{1}+\vec{q}_{2}+\vec{q}_{3}+\vec{q}_{4}=\vec{0})$. Since the field
$\bm\phi$ describes an electric dipole, the action includes a long-range
dipole interaction, and also a coupling to the underlying lattice through the
parameters $r$, $f$, $g$, and $h$. The terms $u$ and $v$ that describe the
local anharmonic interactions give a net positive contribution which ensures
that the polarization remains bounded.
In general these parameters are
tensorial, but for simplicity we have assumed that they adopt cubic
symmetry.
Estimates for the
parameters shown in \tabref{tab:SrTiO3Params} were obtained from \emph{ab
initio} calculations \cite{89pcbmsa04,97lwyk03,03rm01} in the two key
ferroelectrics \SrTiO and \KTaO~\cite{09rssdlss03}.
The typical energy scale of
ferroelectric fluctuations along (100) is $E_{0}=\hbar\pi c/a$; using this
definition we can then employ a dimensionless bosonic Matsubara frequency
$\tilde{\omega}=\omega/E_{0}$, and a dimensionless temperature
$\tilde{T}=T/E_{0}$. Throughout the paper we adopt the units
$a=\hbar=k_{\text{B}}=1$.

\begin{table}
\begin{tabular}{c|ccccccc}
            &$E_{0}/\text{meV}$&$a/\text{\AA}$&$\hbar c/\text{meV}$&$r$  &$f$  &$g$  &$h$ \\
\hline
\SrTiO &4.47              &3.9           &5.55                &5.31 &55.7 &0.39 &5.1 \\
\KTaO  &10.6              &3.9           &13.1                &9.77 &472  &39.2 &165 \\
\end{tabular}
 \caption{Model parameters for the ferroelectrics \SrTiO and \KTaO~\cite{89pcbmsa04,97lwyk03,03rm01,09rssdlss03}.}
 \label{tab:SrTiO3Params}
\end{table}

To establish the connection to previous work we first consider the
mean-field phase diagram that is sketched in
\figref{fig:MeanFieldPhaseDiagram}. Making the ansatz that the ground state
is uniform we obtain the action $S=r\phi^{2}+(u+v)\phi^{4}$, where
$\phi=|{\bm\phi}|$. When $v<0$ the polarization
$\phi_{\text{x}}=\phi_{\text{y}}=0$, $\phi_{\text{z}}^{2}=-r/2(u+v)$ has an
Ising configuration, whereas when $v>0$ the polarization
$\phi_{\text{x}}^{2}=\phi_{\text{y}}^{2}=\phi_{\text{z}}^{2}=-r/2(3u+v)$
exhibits diagonal order. The term proportional to $v$ controls the
polarization direction in the ferroelectric phase, whereas the $u$ term is
rotationally invariant. We note that whilst sweeping $v$ through $v=0$ with
$u>0$ the first order rotation of polarization direction is accompanied with
a continuous change in the magnitude of the polarization. This is driven by
a similar mechanism to the Blume-Emery-Griffiths model involving two bosonic
fields~\cite{71beg09}. Within the mean-field approximation the condition for
stability of the polarization is that the net coefficient of the quartic
term is positive which translates to $u+v>0$ when $v<0$ and $u+v/3>0$ if
$v>0$. If these conditions are not fulfilled then higher order terms must be
included and rather than undergo a second order transition at $r=0$, the
system might have a first order ferroelectric transition at mean-field
level. We can neglect the higher order
terms such as $\lambda\phi^{6}$ provided that the model
remains stable, which requires that $\lambda\phi^{2}\ll u+v$.
Here we wish to investigate whether near criticality the fluctuating modes
can conspire to drive an otherwise second order transition to become first
order. In order to access this behavior we now go beyond mean-field and
consider the consequences of quantum fluctuations on the system.

\section{Field integral formulation}

\begin{figure*}
 \centerline{\resizebox{0.85\linewidth}{!}{\includegraphics{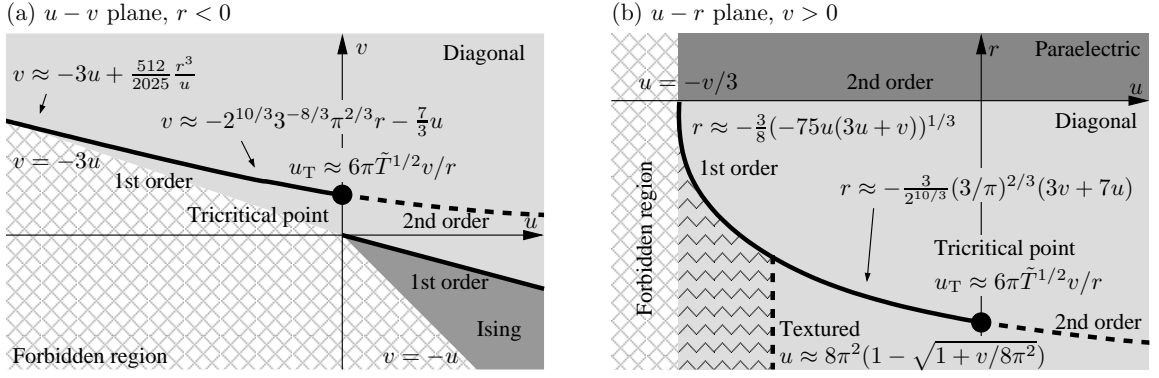}}}
 \caption{The phase diagram at $\tilde{T}=0$ in the (a) $u-v$ plane with
  $r<0$, and (b) $u-r$ plane with $v>0$, both at zero temperature. The
  cross-hatched forbidden region denotes where the polarizability would
  diverge without higher order corrections, the light grey denotes diagonal
  ordered polarization, and the dark grey the (a) Ising phase and (b)
  paraelectric phase. Solid thick lines denote first order phase boundaries,
  dashed lines second order transitions, and the circle the tricritical
  point.}
 \label{fig:PhaseDiagram}
\end{figure*}

To account for fluctuation corrections to the system \citet{03rm01} employed the
renormalization group, which is tailored to study the
well-established second order ferroelectric transition. However, motivated by
recent experiments~\cite{07tiy07,09rssdlss03} we wish to explore the possibility of a first
order metaelectric transition. Therefore, rather than considering just the
corrections due to slow fluctuations that are encompassed by renormalization
group, we need to consider fluctuations $\bm\psi$ over all length scales in the
polarization $\bm\phi+\bm\psi$ around the saddle-point solution $\bm\phi$. When
$u\ll r^{2}$ we can neglect fluctuations in $\bm\psi$ beyond second order which
reduces the action to 
\begin{eqnarray}
 S&=&\tilde{\beta}\left[\left(r+\frac{g}{3}\right)\phi^{2}+u\phi^{4}+v\sum_{\alpha}\phi_{\alpha}^{4}\right]\neweqnline
 &+&\tilde{\beta}\sum_{\tilde{\omega},\vec{q}}{\bm\psi}^{\text{T}}(\tilde{\omega},\vec{q})\matrixsymb{G}^{-1}{\bm\psi}(-\tilde{\omega},-\vec{q})\punc{,}
\end{eqnarray}
where
$\matrixsymb{G}^{-1}_{\alpha,\beta}=G_{\alpha}^{-1}\delta_{\alpha,\beta}+U_{\alpha,\beta}$,
the diagonal inverse Green function takes the form
$G_{\alpha}^{-1}=\tilde{\omega}^{2}+q^{2}+r+fq_{\alpha}^{2}+(g-hq^{2})q_{\alpha}^{2}/q^{2}+(4u+6v)\phi_{\alpha}^{2}+2u\phi^{2}$,
and the off-diagonal terms are
$U_{\alpha,\beta}=(g-hq^{2})q_{\alpha}q_{\beta}/q^{2}+4u\phi_{\alpha}\phi_{\beta}$. We now integrate over quantum
fluctuations to yield the free energy
\begin{equation}
 F=\left(r+\frac{g}{3}\right)\phi^{2}+u\phi^{4}+v\sum_{\alpha}\phi_{\alpha}^{4}+\frac{1}{2\tilde{\beta}}\tr\ln\matrixsymb{G}^{-1}\punc{,}
\end{equation}
where $\tilde{\beta}=1/\tilde{T}$ is the dimensionless inverse
temperature.
If $\bm\phi=\vec{0}$ and $r\gg g-h\pi^{2}$ or if $\bm\phi\ne\vec{0}$ and
$r\ll\pi^{2}$ then $UG\ll1$. In this regime we can expand the inverse Green
function in its off-diagonal terms $U$ using
$\tr\ln\matrixsymb{G}^{-1}=\tr\ln G^{-1}+\tr\ln(1+GU)$ which enables us to
describe the renormalization of fluctuations by off-diagonal coupling. This
yields
\begin{eqnarray}
 \label{eqn:FinalFreeEnergy}
 \!\!&\!\!\!\!\!\!\!\!F&\!\!\!=\!\left(r+\frac{g}{3}\right)\phi^{2}+u\phi^{4}+v\sum_{\alpha}\phi_{\alpha}^{4}\nonumber\\
 \!\!&\!\!\!\!\!\!\!\!+&\!\!\!\frac{1}{\tilde{\beta}}\!\sum_{\alpha}\!\left(\!\tr\ln\sinh\!\left[\!\frac{\tilde{\beta}\xi^{\alpha}_{\vec{q}}}{2}\!\right]\!\!-\!\!\ln\!\left[\!\frac{\tilde{\beta}\xi^{\alpha}_{\vec{0}}}{2}\!\right]\!\right)\!-\!\frac{1}{4\tilde{\beta}}\tr\left(UGUG\right)\!,
\end{eqnarray}
where
$\xi^{\alpha}_{\vec{q}}=[q^{2}+r+fq_{\alpha}^{2}+(4u+6v)\phi_{\alpha}^{2}+2u\phi^{2}]^{1/2}$. To
remove the fluctuations of the static uniform component of ${\bm\psi}$,
which are included in $\phi$, we must introduce the second logarithm. This
has the effect of regularizing the divergence which would otherwise develop
from the first logarithm.  This expression, except for the final fluctuation
correction term, agrees with that of Ref.~\cite{03rm01}, and is analogous to
the coupling of transverse ferromagnetic fluctuations that led the emergence
of first order behavior~\cite{08cs12}. The condition for stability is the
same as for the mean-field case.

The momentum integrals are in general evaluated numerically. However, to
further investigate the diagonal ordered phase we make the approximation
that the cuboid Brillouin zone boundary ($-\pi<q_{\alpha}<\pi$) that bounds
the momentum space integral can be replaced with a spherical boundary that
encloses the same total phase space, so has radius
$\qD=\sqrt[3]{6\pi^{2}}$. In the low temperature limit with the polarization
aligned in the $(1,1,1)$ direction, the resulting integrals can then be
evaluated analytically to yield
\begin{eqnarray}
&\!\!F&\!\!\!=\!\left(r+\frac{g}{3}\right)\phi^{2}+u\phi^{4}+v\sum_{\alpha}\phi_{\alpha}^{4}\nonumber\\
&\!\!+&\!\!\!\frac{3}{32\pi^{2}}\left[\pi\sqrt{\xi+\pi^{2}}\left(\xi+2\pi^{2}\right)\!-\!\xi^{2}\ln\left(\!\!\frac{\pi}{\sqrt{\xi}}+\!\sqrt{1\!+\!\frac{\pi^{2}}{\xi}}\right)\right]\nonumber\\
&\!\!+&\!\!\!\frac{u^{2}\phi^{2}}{6\pi^{2}}\left[\frac{2\pi}{\sqrt{\xi+\pi^{2}}}-2\ln\left(\frac{\pi}{\sqrt{\xi}}+\sqrt{1+\frac{\pi^{2}}{\xi}}\right)\right]\punc{,}
\end{eqnarray}
with $\xi\equiv r+2(u+v)\phi^{2}+4u\phi^{2}/3$, and were found to be in good
agreement with the corresponding numerical result.

\subsection{Phase behavior and heat capacity}

The phase behavior of the system is shown in \figref{fig:PhaseDiagram}. The
forbidden region indicates where the action polarizability and free energy
would diverge without considering higher order corrections to the original
action. When considered within the framework of mean-field phenomenology,
here the system could undergo a first order paraelectric-ferroelectric
transition. However, the corrections due to quantum fluctuations renormalize
the action, causing a metaelectric boundary to peel away from the first
order transition associated with the forbidden region. This metaelectric
transition is consistent with recent experimental evidence for a first order
phase transition~\cite{07tiy07} in $^{18}$O-exchanged SrTiO$_{3}$. In both of the
planes considered, the line of first order metaelectric transitions covers
an extensive region of the phase diagram, terminating in a tricritical point
at $u=0$.
The first order transition at small $u$ is destroyed at non-zero
temperature, with the tricritical point moving up the line of transitions to
$u\approx6\pi\tilde{T}^{1/2}v/r$. This critical behavior does not depend on
the long-range dipole interactions since the lowest order term in $g$ and
$h$ averages to zero on integrating over momenta~\cite{03rm01}. The
$\phi\rightarrow-\phi$ symmetry could be destroyed by applying a uniaxial
electric field misaligned to the lattice.


A further ramification of the quantum fluctuation corrections is that the
rotation of the polarization from Ising to diagonal order no longer occurs
where $v$ turns negative. Though, as for the mean-field case, the magnitude
of the polarization is conserved; fluctuations have renormalized the quartic
terms and shifted the phase boundary in \figref{fig:PhaseDiagram}(b). This
behavior can also be recovered by a renormalization group analysis
\cite{03rm01}. One experimental probe of the metaelectric transition is the
changing behavior of the heat capacity $C=-T\partial^{2}F/\partial
T^{2}$. Before the metaelectric transition (small negative $r$) the relevant
optic mode is ``soft'' and so the heat capacity follows the familiar Debye
form $C\sim T^{3}$, whereas after the metaelectric transition (large
negative $r$), the relevant optic modes are ``stiff'' and so the heat
capacity has an exponential dependence on temperature. At high temperature,
in both cases the heat capacity has the expected classical behavior
$C=3k_{\text{B}}$.

Having confirmed the existence of a possible metaelectric behavior, we now
turn to consider the stability of the phase in the vicinity of the
transition.  Recent studies of itinerant ferromagnetism have suggested that
such first order behavior can be preempted by the development of textured
magnetic order analogous to that seen in the FFLO phase of
superconductors~\cite{09cgs06}. This leaves open the question as to whether
a textured phase can develop in the vicinity of the metaelectric
transition. Our strategy to explore this possibility is to assume that the
inhomogeneous phase is formed continuously, which allows us to develop a
Landau expansion in the polarization $\Phi$ and texture wave vector
$\vec{Q}$. The onset of an inhomogeneous phase is signaled by the
coefficient of the $\Phi^{2}Q^{2}$ term turning negative. In our analysis we
search primarily in the vicinity of the metaelectric transition at $\xi=0$
and consider a trial state with uniform polarization $\phi$ that is for
simplicity superimposed by an inhomogeneous component
$\Phi\cos(\vec{Q}\cdot\vec{r})(1,1,1)$. We then expand the free energy to
quartic order in $\vec{Q}$ and discover that the presence of a textured
phase makes a contribution to the total energy of
$Q^{2}\Phi^{2}[1-u^{2}\Phi^{2}/6\pi^{2}\xi+7u^{2}\Phi^{2}Q^{2}/60\pi^{2}\xi^{2}]$.
Short of the first order transition where $\xi<0$, the coefficient of
$Q^{2}$ is positive so the phase is not modulated. After the first order
transition $\xi$ turns positive driving the coefficient of $Q^{2}$ negative,
revealing a finite $Q$ instability in the region highlighted
in~\figref{fig:PhaseDiagram}(b). The modulation carries polarization
$\Phi=r/2(u+3/v)$. Though the analysis is restricted to the consideration of
a potential continuous transition into the textured phase, and
a simple form for the texture, it is sufficient to validate its
existence. Refinements to include a putative first order transition
or further textured phases would only enlarge the region of the phase diagram
over which inhomogeneities could be observed. Leaving aside potential
textured phases we now turn to consider the behavior of the susceptibility
across the phase diagram.

\subsection{Inverse susceptibility}

The inverse susceptibility provides an experimental
window~\cite{80rhb07,06c03,09rssdlss03} onto the quantum critical properties
of ferroelectrics. Deep in the paraelectric regime
where $R\equiv r+g/3\gg q_{\text{D}}$, the contribution
to the inverse susceptibility is
$\chi^{-1}=\partial^{2}F/\partial\phi^{2}|_{\phi_{\text{eqm}}}=
R+\frac{5u+3v}{\pi^{2}}+\frac{R}{6}(\gamma-\tan^{-1}\gamma)\coth(\frac{\sqrt{R}}{2\tilde{T}})$,
which is consistent with Barrett's formula
\cite{52b04} for a gapped system.
In the quantum critical regime we see
three characteristic types of behavior for the inverse susceptibility
\begin{eqnarray*}
 \!\!\!\!\chi^{-1}\simeq R&&\nonumber\\
  \!\!\!\!+\frac{5u+3v}{\pi^{2}}&&\!\!\!\!\!\!\begin{cases}
  \frac{R}{4}(\gamma\sqrt{1+\gamma^{2}}-\sinh^{-1}\gamma)+\frac{\pi^{2}\tilde{T}^{2}}{18}&\tilde{T}\ll\frac{\qD}{2}\\
  \frac{\sqrt{R}}{3}(\gamma-\tan^{-1}\gamma)\tilde{T}&\tilde{T}\gg\frac{\qD}{2}
 \end{cases}\nonumber\\
  +&&\!\!\!\!\!\!\begin{cases}0&\;\;\;\;\;\;\;\;\;\;\;\;\;\;\;\;\;\;\;\;\;\;\tilde{T}\ll\frac{\sqrt{g}}{2}\\
  \frac{5u+3v}{20}(\frac{g}{6}-\sqrt{g}h)\tilde{T}&\;\;\;\;\;\;\;\;\;\;\;\;\;\;\;\;\;\;\;\;\;\;\;\tilde{T}\gg\frac{\sqrt{g}}{2}
  \end{cases}\nonumber\\
  +\frac{h^{2}}{15\pi}&&\!\!\!\!\!\!\begin{cases}(5u+3v)(\frac{3\qD^{2}}{16}+\frac{\pi^{2}}{2}\tilde{T}^{2})&\;\;\;\;\;\;\;\;\;\;\;\;\;\;\;\tilde{T}\ll\frac{q_{\text{D}}}{2}\\
  2\qD\tilde{T}&\;\;\;\;\;\;\;\;\;\;\;\;\;\;\;\tilde{T}\gg\frac{q_{\text{D}}}{2}
 \end{cases}\punc{,}
\end{eqnarray*}
where the first term is from the diagonal contributions to
\eqnref{eqn:FinalFreeEnergy}, and the latter two terms are the off-diagonal
contribution, and $\gamma=\qD/R$. At low temperature effects of
long-range dipole interactions prevail as the off-diagonal fluctuating
contribution renormalizes the on-diagonal terms, with the linear temperature
dependence of the term proportional to $g$ giving a positive slope to the
inverse susceptibility whereas the $\sqrt{g}h$ term could provide a negative
slope. At higher temperatures the $T^{2}$ contribution from the mean-field
term dominates, which is also characteristic of quantum critical behavior
and is in good agreement with recent experimental
results~\cite{09rssdlss03}. We note that the $T^{2}$ behavior is recovered
by other models, including a diagrammatic resummation~\cite{71r08,71ks09},
the quantum spherical model~\cite{76sbs02}, renormalization group
studies~\cite{83s07,97s01}, a self-consistent phonon
model~\cite{09rssdlss03}, and an analogy to the temporal Casimir
effect~\cite{09pcc02}. The behavior has also been observed
experimentally~\cite{80rhb07,06c03,09rssdlss03}. In both \SrTiO and \KTaO
the initial linear negative slope and the quadratic $\chi^{-1}\sim T^{2}$
term conspire to cause a characteristic minimum in the inverse
susceptibility.  Using estimates for the parameters
in \tabref{tab:SrTiO3Params}, the minimum occurs at
$T\sim1\unit{K}$ in both \SrTiO and \KTaO which is in good agreement with
the experimental values of $T=1.6\unit{K}$ and $T=3.0\unit{K}$
respectively~\cite{09rssdlss03}. Finally, at high temperatures a classical
term $\chi^{-1}\sim T$ from the longitudinal fluctuating term dominates from
$\sim100\text{K}$ which is again in good agreement with the experimental
observations~\cite{09rssdlss03}.

\section{Discussion}

In this paper we have found that the polar fluctuating phonons can drive a
displacive ferroelectric through a first order metaelectric transition. Long
range dipolar interactions did not affect this critical phase
behavior~\cite{03rm01}. However, long-range dipole interactions introduced
into the action through the term $(g-hq^{2})\phi^{2}$ were pivotal in
creating the correction to the inverse susceptibility $\chi^{-1}\sim-T$ that
could explain the characteristic inverse susceptibility
minimum~\cite{09rssdlss03}, as well as provide important corrections to the
self-consistent phonon treatment~\cite{09rssdlss03}.

However, another mechanism, coupling of the soft optic modes to acoustic
phonons could be significant. It has already been
understood~\cite{71ks09,09pcc02,09rssdlss03} that a coupling with the
acoustic phonons $\varphi$ of the form $-\eta(\nabla\varphi)\phi^{2}$ leads
to a correction in $\chi^{-1}$ of $-T^{4}$ that could explain the
characteristic minimum in the inverse susceptibility, and also has the
capability of driving a first order
transition~\cite{09pcc02,09rssdlss03}. This work and the results presented
here motivate further experimental investigations into the inverse
susceptibility and putative metaelectric transition that could shed light on the
origin of the phase structure. Though the coupling to acoustic phonons
complicated the solid state system, ultracold atoms in an optical lattice
with long-range dipole interactions~\cite{02bdgsl01} present a clean system
that could provide a powerful tools to help unravel the properties of the
generic Hamiltonian.

One important simplification was to model the ferroelectric with undamped
dynamics. Damping would primarily arise due to free electrons, which can be
introduced controllably through doping. Analogous to ``avoided criticality''
at a magnetic critical point which leads to non-Fermi liquid behavior and
superconductivity, ferroelectrics might also be expected to adopt novel
behavior; for example doped \SrTiO~\cite{65shabck03}, whereas undoped
SnTe~\cite{76kkkk09} and GeTe~\cite{06bmam06} become superconducting at low
temperatures. This area presents a promising avenue of research. Further
open questions are to determine whether with just a change of parameters
\cite{01rj04,03rm01} the same formalism be applied to order-disorder
ferroelectrics, and to consider the consequences of the coupling of
fluctuating polarization and magnetization that could arise in
\EuTiO~\cite{08ssrkpsp08}.

We thank Mark Dean, Gil Lonzarich, Stephen Rowley, and Montu Saxena for useful discussions.


\end{document}